\documentclass[a4paper,12pt]{article}

\usepackage{amsmath,amssymb,graphicx}
\usepackage[active]{srcltx}
\usepackage{cite}
\usepackage{color}
\numberwithin{equation}{section}

\newcommand{\nn}{\nonumber}

\newcommand{\ii}{\mathrm{i}}

\newcommand{\dd}{\mathrm{d}}

\newcommand{\La}{\mathcal{L}}

\newcommand{\e}{\mathrm{e}}

\newcommand{\tr}{\mathop{\mathrm{tr}}\nolimits}

\newcommand{\N}{\mathcal{N}}

\begin{document}

\baselineskip=.22in

\title{Enhanced Supersymmetry of Nonrelativistic ABJM Theory
}%
\author{O-Kab Kwon$^{1),}$\thanks{email: okab@skku.edu},
\and Phillial Oh$^{1),}$\thanks{email: ploh@skku.edu},
\and Corneliu Sochichiu$^{2),}$\thanks{email: sochichi@skku.edu},
\and Jongsu Sohn$^{1),}$\thanks{email: jongsusohn@skku.edu} \and \\
$^{1)}${\it Department of Physics,~BK21 Physics Research Division,}\\
{\it Institute of Basic Science, Sungkyunkwan University, Suwon 440-746, Korea}\\
$^{2)}${\it University College,
Sungkyunkwan University, Suwon 440-746, Korea}\\
and \\
{\it Institutul de Fizic\u a Aplicat\u a A\c S,}\\
 {\it str. Academiei, nr. 5, Chi\c{s}in\u{a}u, MD2028 Moldova.
}
}
%

\maketitle
\begin{abstract}
We study the supersymmetry enhancement of nonrelativistic limits
of the ABJM theory for Chern-Simons level $k=1,2$. The special
attention is paid to the nonrelativistic limit (known as `PAAP'
case) containing both particles and antiparticles. Using
supersymmetry transformations generated by the monopole operators,
we find additional  2 kinematical, 2 dynamical, and 2 conformal
supercharges for this case. Combining with the original 8
kinematical supercharges, the total number of supercharges becomes
maximal: 14 supercharges, like in the well-known PPPP limit. We
obtain the corresponding super Schr\"odinger algebra which appears
to be isomorphic to the one of the PPPP case. We also discuss the
role of monopole operators in supersymmetry  enhancement and
partial breaking of supersymmetry in nonrelativistic limit of the
ABJM theory.

\end{abstract}
\maketitle

\newpage

\section{Introduction}

In the shed of recent development of M2-brane theories the three
dimensional Chern-Simons based superconformal
theories~\cite{Schwarz:2004yj} gained a considerable interest. Two
different types of theories were proposed as candidates to
effectively describe multiple M2-branes system: the 3-algebra
based BLG (Bagger--Lambert and Gustavsson)
theory~\cite{Bagger:2006sk,Gustavsson:2007vu} and standard Lie
algebra based ABJM (Aharony--Bergman--Jafferis--Maldacena)
theory~\cite{Aharony:2008ug}. The former theory
is restricted to describe the effective action of two M2-branes only.
As a result of this, lately the main interest focused on the ABJM
and related theories, in which the setup with an arbitrary number
of M2-branes is allowed. Beyond that, the ABJM theory proved to reproduce
the correct vacuum structure of M2-branes
on ${\mathbb C}^4/{\mathbb Z}_k$-orbifold, where $k$ appears
as the Chern-Simons level in the theory.

The BLG theory, which is equivalent to the ABJM theory with
SU(2)$\times$SU(2) gauge group~\cite{Aharony:2008ug}, has ${\cal N}=8$
supersymmetry (SUSY).
On the other hand, the original ABJM theory has only manifest ${\cal N}=6$
supersymmetry instead of $\N=8$ expected from the M2-brane system
in flat transverse space.
The lack of supersymmetries can be explained by the fact that for
the Chern-Simons level $k$ the theory describes multiple M2-branes
in ${\mathbb C}^4/{\mathbb Z}_k$-orbifold rather than in flat space
and $\mathcal{N}=6$ is the maximal supersymmetry which the
${\mathbb C}^4/{\mathbb Z}_k$-orbifold allows, if $k\geq 3$.
In what concerns the cases of $k=1,2$, it was conjectured from the dual
gravity that the full ${\cal N}=8$ supersymmetry should be
restored~\cite{Aharony:2008ug}.
Indeed, using \emph{monopole operators}, it was proved that the ABJM
theory has additional ${\cal N}=2$ supersymmetry for
$k=1,2$~\cite{Gustavsson:2009pm,Kwon:2009ar}.
At these levels, attaching a monopole operator to a local field changes
the rule of gauge transformation of the field, while preserving
the local nature
of the field~\footnote{For the detailed study of monopole operators in ABJM
theory and related topics,
see Refs.~\cite{Berenstein:2008dc,Klebanov:2008vq,Park:2008bk,
Imamura:2009ur,Kim:2009wb,SheikhJabbari:2009kr,Berenstein:2009sa,Benna:2009xd,
Kim:2009ia,Imamura:2009hc}.}.
Using this procedure we can introduce new fields in the theory with the
help of monopole operators.
In Ref.~\cite{Gustavsson:2009pm}  the additional ${\cal N}=2$ supersymmetry
and SO(8)-invariance of the ABJM potential were obtained for
U($N$)$\times$U($N$)
gauge group at $k=1,2$ in 3-algebra formulation.
The same enhancement of supersymmetry in the Lie-algebra formulation of
ABJM theory for the case of U(2)$\times$U(2) gauge group was found in
Ref.~\cite{Kwon:2009ar}, where the so called \emph{minimal model}
which inherits most properties of the ABJM theory was introduced.
This minimal model can be used as a toy model replacement for the ABJM theory.

Recently there has been much interest in nonrelativistic
versions of AdS/CFT correspondence~\cite{Son:2008ye,Balasubramanian:2008dm,
Herzog:2008wg,Goldberger:2008vg,Barbon:2008bg,Maldacena:2008wh,Adams:2008wt,
Hartnoll:2008rs,Donos:2009en,Volovich:2009yh,Bobev:2009mw,Donos:2009zf}.
The geometrical framework for nonrelativistic (super)symmetry was studied
earlier in \cite{Duval:1990hj} (see also \cite{Duval:2008jg}).
Following the development
of superconformal field theories describing the dynamics of M2-branes,
the construction of nonrelativistic superconformal field theories and finding
their gravity duals become of increasing interest.
In the past year, a number of nonrelativistic limits of the ABJM theory
were constructed~\cite{Nakayama:2009cz,Lee:2009mm}.
The nonrelativistic AdS/CFT
correspondence~\cite{Colgain:2009wm,Ooguri:2009cv,Nakayama:2009ed,Jeong:2009aa}
in relation with M-theory and
the soliton solution~\cite{Kawai:2009rc} in nonrelativistic ABJM theory
were also studied.

It is known that some nonrelativistic limits result in additional symmetries,
which are not present in the original theory.
Schr\"odinger symmetry~\cite{Hagen:1972pd} is such an example.
Emergence of new bosonic symmetries can be accompanied by the emergence of
new fermionic ones.
However, there are not so many examples where the explicit construction
of nonrelativistic superconformal field theories and
analysis of their super Schr\"odinger algebras were
done~\cite{Leblanc:1992wu,Nakayama:2008qm,Nakayama:2008qz,Nakayama:2008td}.

In this paper we find additional 6 supercharges using the
\emph{monopole operators} and complete the corresponding super Schr\"odinger
algebra in a nonrelativistic limit of the ABJM theory,
referred in Ref.~\cite{Nakayama:2009cz} as
`PAAP limit'\footnote{See the subsection \ref{nrlims} for the definitions of various
nonrelativistic regimes.}.

In general, the nonrelativistic limit of a field theory is ambiguous
and depends on the particle-antiparticle sectors chosen.
The supersymmetry of various nonrelativistic limits of the ABJM theory was
studied by several authors in~\cite{Nakayama:2009cz,Lee:2009mm}.
These authors studied supersymmetry and super Schr\"odinger algebras
for different non-relativistic regimes.
In particular, it was found that one can define non-relativistic limits
in which different numbers of supersymmetries are
conserved~\cite{Nakayama:2009cz}.
The `basic' ${\cal N}=6$ supersymmetry in the mass deformed ABJM
theory~\cite{Hosomichi:2008jb,Gomis:2008vc} survives and resulting
12 supercharges are split into 10 kinematical and  2 dynamical ones
in the limit which involves only particles (PPPP limit).
Additional 2 supercharges emerge coming from conformal symmetry
associated with different scalings of time and space.
However, if we include antiparticle sectors,
the number of supersymmetry is reduced and superconformal symmetry is
broken~\cite{Nakayama:2009cz}. This is considered as a general property of
the nonrelativistic limits of superconformal field
theories~\cite{Nakayama:2008qz,Nakayama:2008td}.

Our aim in this work is to extend the analysis to supersymmetry
driven by \emph{monopole operators}.
In nonrelativistic limit where both particles and antiparticles are
included e.g. PAAP limit, it appears that some of the supersymmetry
can be restored. In order to have the enhanced supersymmetry,
we should restrict ourself to the Chern-Simons levels $k=1,2$.
Although the most analysis is done for the gauge group U(2)$\times$U(2),
we believe our conclusions hold true for the general case of
U($N$)$\times$U($N$).

The plan of the paper is as follows.
In the next section we review the mass deformation and nonrelativistic limits
of the ABJM theory following~\cite{Nakayama:2009cz}.
In the third section we introduce the PAAP limit and describe the known
8 kinematical supercharges as well as bosonic symmetries.
In the fourth section we analyze the enhanced symmetry generated by
monopole operator. We find that in the PPPP limit
with maximal basic supersymmetry $Q_1+Q_2+S=10+2+2$,
the monopole enhanced part of supersymmetry~\cite{Gustavsson:2009pm,Kwon:2009ar}
is broken\footnote{ We denote the numbers of kinematical, dynamical,
and conformal supercharges as $Q_1$, $Q_2$, and $S$ respectively.}.
In contrast, in the PAAP limit, where the basic nonrelativistic symmetry
is broken down to $Q_1+Q_2+S=8+0+0$~\cite{Nakayama:2009cz},
the enhanced supersymmetry gives rise to $2+2+2$ new supercharges
including conformal ones. The total number of supercharges is the same as
in the PPPP limit. We also derive the explicit form of these supercharges and
their (anti)commutation relations to obtain the maximal super
Schr\"odinger algebra.
The last section is our conclusion.

\section{Mass-deformed ABJM Theory}
The ABJM theory~\cite{Aharony:2008ug} was proposed as a candidate to describe
multiple M2-brane systems. The model is a three-dimensional gauge theory
which possesses $\mathcal{N}=6$ supersymmetry.
The field content of this model is given by four bi-fundamental complex scalars
$Y^A$, $A=1,\dots,4$, four bi-fundamental complex spinors $\Psi_A$ as well as two
Chern-Simons gauge fields $A_{\mu}$ and $\hat{A}_\mu$, $\mu=0,1,2$.
The scalars and the fermions
belong to the four dimensional complex representations of $R$-symmetry
group SU(4).
The SU(4) indices can be conveniently written in terms of more handy
SU(2)$_{{\rm L}}\times$SU(2)$_{{\rm R}}$ notations:
$A=(n,n')$, where $n$ and $n'$ are SU(2)$_{{\rm L}}$ and SU(2)$_{{\rm R}}$
indices respectively.
This decomposition is even more natural in the mass-deformed ABJM theory
which has SU(2)$_{{\rm L}}$$\times$SU(2)$_{{\rm R}}$$\times$U(1)$_R$ symmetry.
Restriction to the sector of a single SU(2) factor gives the minimal
model~\cite{Kwon:2009ar}. For SU(2) indices we will use the Latin $m$
and $n$ for the SU(2)$_{{\rm L}}$ as well as primed letters for the SU(2)$_{{\rm R}}$.

As about spinor notations, we choose
(2+1)-dimensional gamma matrices which satisfy
$\gamma^\mu\gamma^\nu=\eta^{\mu\nu} +
\epsilon^{\mu\nu\rho}\gamma_\rho$ as $ \gamma^0= i\sigma^2,
\gamma^1=\sigma^1$, and $\gamma^2=\sigma^3$. The suppressed spinor
indices are expressed by $\xi \chi\equiv \xi^\alpha\chi_\alpha$ and
$\xi\gamma^\mu\chi=\xi^\alpha\gamma_{\alpha}^{\mu\,\beta}\chi_\beta$
for the two component spinors $\xi$ and $\chi$.
For the gauge indices  we use the convention of the Ref.~\cite{Benna:2008zy}.

\subsection{Mass deformation}
The SUSY preserving mass deformation was first constructed for the
BLG theory~\cite{Gomis:2008cv,Hosomichi:2008qk}. Then it was found
that the ABJM theory also admits the SUSY preserving mass
deformation~\cite{Hosomichi:2008jb,Gomis:2008vc}.
There are several methods to obtain the mass-deformed ABJM theory,
such as ${\cal N}=1$ superfield formalism~\cite{Hosomichi:2008qk},
$D$-term and $F$-term deformations~\cite{Gomis:2008vc}
in ${\cal N}=2$ superfield formalism~\cite{Benna:2008zy}.
These different versions of mass-deformed ABJM theory are equivalent~\cite{Kim:2009ny}
and the M-theory origin of the mass-deformation
was investigated in Refs.~\cite{Kim:2009nc,Lambert:2009qw}.
Below we consider the mass deformed version of ABJM theory in more details.

The mass deformed ABJM theory with U($N$)$\times$U($N$) gauge group
is given by the action~\cite{Hosomichi:2008jb,Gomis:2008vc},
\begin{align}\label{ABJMact}
S =c\int dtd^2x\,\left({\cal L}_0 + {\cal L}_{{\rm CS}} -V_{{\rm ferm}}
-V_{{\rm bos}} -V_{{\rm m}}\right),
\end{align}
where
\begin{subequations}
\begin{align}
{\cal L}_0 &= {\rm tr}\left(\frac{1}{c^2}
D_t Y_A^\dagger D_tY^A-D_i Y_A^\dagger D_i Y^A +
\frac{i}{c}\Psi^{\dagger A} \gamma^0 D_t \Psi_A+i\Psi^{\dagger A}
\gamma^iD_i\Psi_A\right),
\label{ABJMkin} \\
{\cal L}_{{\rm CS}} &= \frac{k\hbar}{4\pi}\,\epsilon^{\mu\nu\rho}\,{\rm tr}
\left(A_\mu \partial_\nu A_\rho +\frac{2i}{3}A_\mu A_\nu A_\rho
- \hat{A}_\mu \partial_\nu \hat{A}_\rho
-\frac{2i}{3}\hat{A}_\mu \hat{A}_\nu \hat{A}_\rho \right),
\label{ABJMcs} \\
V_{{\rm ferm}} &= \frac{2\pi i}{k\hbar c}{\rm tr}
\Big( Y_A^\dagger Y^A\Psi^{\dagger B}\Psi_B
-Y^A Y_A^\dagger\Psi_B \Psi^{\dagger B}
+2Y^AY_B^\dagger\Psi_A\Psi^{\dagger B} -2Y_A^\dagger
Y^B\Psi^{\dagger A}\Psi_B
\label{ferV} \\
&\hskip 1.7cm  +\epsilon^{ABCD}Y^\dagger_A\Psi_BY^\dagger_C\Psi_D
-\epsilon_{ABCD}Y^A\Psi^{\dagger B}Y^C\Psi^{\dagger D} \Big),
\nn \\
V_{{\rm bos}} &=-\frac{4\pi^2}{3k^2\hbar^2 c^2}{\rm tr}\Big(
Y^\dagger_AY^AY^\dagger_BY^BY^\dagger_CY^C
+Y^AY^\dagger_AY^BY^\dagger_BY^CY^\dagger_C
+4Y^\dagger_AY^BY^\dagger_CY^AY^\dagger_BY^C
\nn \\
&\hskip 2.8cm -6Y^AY^\dagger_BY^BY^\dagger_AY^CY^\dagger_C \Big),
\label{bosV} \\
V_{{\rm m}}&={\rm tr}\left[ \frac{imc}{\hbar} M_{A}^{~B}\Psi^{\dagger A}\Psi_B
+\frac{4\pi m}{k\hbar^2}M_{B}^{~C}\left(Y^AY_A^\dagger Y^B Y_C^\dagger
-Y_A^\dagger Y^A Y_C^\dagger Y^B\right)+\frac{m^2c^2}{\hbar^2}
Y_A^\dagger Y^A\right],\label{dpot}
\end{align}
\end{subequations}
and the explicit dependence on the speed of light $c$ and Planck constant
$\hbar$ is given.

The last term \eqref{dpot} in the action \eqref{ABJMact} represents the
mass deformation with $m$ being the mass parameter and
$M_A^{~B}={\rm diag}(1,1,-1,-1)$. Without this term equation
\eqref{ABJMact} represents the original \emph{massless} ABJM action.
The covariant derivatives are defined as
\begin{align}\label{covD2}
D_t Y^A &=\partial_t Y^A +iA_t Y^A -iY^A \hat{A}_t,\,\,
A_t = cA_0={\rm finite},
\nn \\
D_i Y^A &=\partial_i Y^A +i A_i Y^A-iY^\dagger_A \hat A_i.
\end{align}

\subsection{${\cal N}=6$ Supersymmetry}
The action \eqref{ABJMact} of the \emph{massless} ABJM theory is invariant
with respect to the following $\mathcal{N}=6$ supersymmetry transformations,
\begin{align}
&\delta Y^A = \ii \omega^{AB}\Psi_B, \qquad
\delta Y^{\dagger}_A = \ii \Psi^{\dagger B}\omega_{AB},
\nn \\
&\delta\Psi_A = \gamma^\mu\omega_{AB}D_\mu Y^B
+\frac{2\pi}k\omega_{AB}(Y^BY^\dagger_CY^C -Y^CY^\dagger_CY^B)
+\frac{4\pi}k\omega_{BC}Y^BY^\dagger_AY^C,
\nn \\
&\delta\Psi^{\dagger\,A}= -D_\mu Y^\dagger_B \omega^{AB}\gamma^\mu
+\frac{2\pi}k\omega^{AB}(Y^\dagger_CY^CY^\dagger_B
-Y^\dagger_BY^CY^\dagger_C)
-\frac{4\pi}k\omega^{BC}Y^\dagger_BY^AY^\dagger_C,
\nn \\
&\delta A_\mu = -\frac{2\pi}{k}(\omega^{AB}Y^{\dagger}_A
\gamma_\mu\Psi_B + Y^A \Psi^{\dagger B}\gamma_\mu\omega_{AB}),
\nn \\
&\delta \hat{A}_\mu =
-\frac{2\pi}{k}(\omega^{AB}Y^\dagger_A\gamma_\mu\Psi_B
+\Psi^{\dagger B}\gamma_\mu Y^A\omega_{AB}),
\label{N6part}
\end{align}
where $\omega^{AB}=-\omega^{BA}=(\omega_{AB})^*=
\frac{1}{2}\,\epsilon^{ABCD}\omega_{CD}$. The supersymmetric
parameters $\omega_{AB}$ and $\omega^{AB}$ are related to
the (2+1)-dimensional six Majorana spinors $\varepsilon_I$,
$I=1,\cdots,6$, by
\begin{align}\label{omegam}
\omega_{AB} = \varepsilon_{I}(\Gamma^I)_{AB},
\qquad
\omega^{AB} = \varepsilon_{I}(\Gamma^{I*})^{AB}.
\end{align}

As we mentioned above, the mass deformation preserves the full
$\mathcal{N}=6$ supersymmetry, the only effect of such deformation
being the modification of the spinor field transformation rules by
the additional terms,
\begin{align}
\delta_m \Psi_A &= m M_A^{~B} \omega_{BC} Y^C,
\nonumber \\
\delta_m\Psi^{\dagger A} &= m M^A_{~B} \omega^{BC} Y_C^\dagger.
\end{align}

\subsection{Nonrelativistic limit(s)}\label{nrlims}
Formally, nonrelativistic limit corresponds to the limit of large
speed of light $c\to\infty$. However, this limit is not uniquely defined.
Below we give a brief description of nonrelativistic limits of our interests
in ABJM theory.

As was discussed before,
the mass deformation breaks the original SU(4) $R$-symmetry
down to SU(2)$_{{\rm L}}$$\times$SU(2)$_{{\rm R}}$$\times$U(1)$_R$ $R$-symmetry.
According to this the fields are split as,
\begin{align}
Y^A &= (Y^n, Y^{n'}),\quad Y_A^\dagger = (Y_n^\dagger, Y_{n'}^\dagger),
\nn \\
\Psi_A &= (\Psi_n, \Psi_{n'}), \quad
\Psi^{\dagger A} = (\Psi^{\dagger n}, \Psi^{\dagger n'}),
\end{align}
where $A=1,2,3,4$, $n=1,2$, and $n'=3,4$.

In order to go to the nonrelativistic limit,
we decompose the relativistic fields into the particle
and antiparticle parts,
\begin{equation}\label{decom}
Y^n =\frac{\hbar}{\sqrt{2m}}\Big(\e^{-\ii mc^2t/\hbar} y^n
+ \e^{\ii mc^2t/\hbar} \hat y^{\dagger n}\Big),\quad
\Psi_n =\sqrt{\hbar c}\Big(\e^{-\ii mc^2t/\hbar}\psi_n +
\e^{\ii mc^2t/\hbar}\sigma_2\hat\psi^\dagger_n\Big),
\end{equation}
and analogously for $Y^{n'}$ and $\Psi_{n'}$. Here the minus sign
in the exponent $\e^{\pm\ii mc^2t/\hbar }$ corresponds to
a particle, while plus sign comes with an antiparticle.

In this situation we can make a choice for each of complex fields
$(Y^n,Y^{n'},\Psi_{n},\Psi_{n'})$ to be either particle (P), or
antiparticle (A) separately. Following~\cite{Nakayama:2009cz},
we will denote such a choice by a four letter string consisting of
`P's and `A's, e.g. the limit with all fields chosen to be in purely
particle sector is denoted PPPP; the limit in which $Y^{n}$ and
$\Psi_{n'}$ are particles while $Y^{n'}$ and $\Psi_{n}$ antiparticles
is denoted PAAP, etc.

To obtain the nonrelativistic limit we have to plug the Ansatz
\eqref{decom} (where the choice is made for the particles or antiparticles)
into the mass-deformed action (\ref{ABJMact}),
take the limit $c\to\infty$, keeping the leading terms.
In the case of fermions we have to eliminate the heavy modes
using their equations of motion~\cite{Nakayama:2009cz,Lee:2009mm}.

\section{PAAP Limit}

The basic supersymmetry in various nonrelativistic limits was
studied in~\cite{Nakayama:2009cz}, where the number of supersymmetries
was found for each limit labeled by the four-letter string. In general,
the number of supercharges depends on the relative choice of fields
as particles or antiparticles. The maximal supersymmetry is
reached in the case of either PPPP, when all fields are particles
or AAAA, when all are antiparticles. In this case the theory possesses
$14$ supercharges of which 10 supercharges are kinematical,
2 are dynamical, and the remaining 2 are conformal related to super Schr\"odinger
symmetry. Other cases contain less supersymmetry of this kind~\cite{Nakayama:2009cz}.
That is, if we include antiparticle sectors in nonrelativistic limits of the
ABJM theory, the number of supercharges is reduced. This is claimed to be
a general property in the nonrelativistic limits of superconformal field
theories~\cite{Nakayama:2008qz,Nakayama:2008td}. For instance,
the PAAP case, which we are going to consider in more details below,
possesses only eight kinematical supercharges and no conformal supercharges
are present, i.e. $Q_1+Q_2+S=8+0+0$.

As we know about the SUSY enhancement in relativistic ABJM theory,
we have to use the monopole operators to obtain the additional
${\cal N}=2$ supersymmetry~\cite{Gustavsson:2009pm,Kwon:2009ar}.
By using monopole operators,
we can include both {\bf 4} and ${\bf {\bar 4}}$ matter fields
simultaneously in ${\cal N}=8$ supersymmetry transformation rules for $k=1,2$.
In the nonrelativistic limit,
however, we cannot do this, since we have
to choose between particle or antiparticle parts for a given relativistic field.
(See the subsection \ref{mono} for the detailed explanations.)
Therefore the ${\cal N}=8$ supersymmetry in relativistic ABJM theory for
$k=1,2$ is inevitably broken in any nonrelativistic limit.

So far, the most supersymmetric case is the PPPP limit with 14 supercharges.
However, since the monopole operators generate supersymmetry transformations
which mix  the {\bf 4} and ${\bf {\bar 4}}$ representations,
we can expect other limits with maximal supersymmetry,
which will include antiparticle sectors.
Indeed, we find that the PAAP limit is another case which has 14 supercharges using
monopole operators.

\subsection{Action}
Let us consider the PAAP case in more details. In PAAP limit one
chooses the first boson and second fermion to be in particle sector
while second boson and first fermion to be antiparticles,
\begin{align}\label{PAAP}
Y^n &=\frac{\hbar}{\sqrt{2m}}\, e^{-imc^2 t/\hbar} y^n,\quad
Y_n^\dagger = \frac{\hbar}{\sqrt{2m}}\, e^{imc^2t/\hbar}y_n^\dagger,
\nn \\
Y^{n'} &=\frac{\hbar}{\sqrt{2m}}\, e^{imc^2 t/\hbar}
\hat y^{\dagger n'},\quad
Y_{n'}^\dagger = \frac{\hbar}{\sqrt{2m}}\,
e^{-imc^2t/\hbar}\hat y_{n'},
\nn \\
\Psi_n &= \sqrt{\hbar c}\, e^{imc^2t/\hbar}\sigma_2\hat\psi_n^\dagger,
\quad
\Psi^{\dagger n} = -\sqrt{\hbar c}\,
e^{-imc^2t/\hbar}\sigma_2\hat\psi^n,
\nn \\
\Psi_{n'} &= \sqrt{\hbar c}\, e^{-imc^2t/\hbar}\psi_{n'},
\quad
\Psi^{\dagger n'} = \sqrt{\hbar c}\, e^{imc^2t/\hbar}\psi^{\dagger n'}.
\end{align}

Applying the general procedure for nonrelativistic limit mentioned before,
we obtain the following nonrelativistic action,
\begin{equation}\label{PAAP-action}
  S_{\rm PAAP}=\int\dd t\dd^2x(\La_{\rm scalar}+\La_{\rm fermion}
  +\La_{CS}),
\end{equation}
where the scalar and fermionic parts of the Lagrangian $\La_{\rm scalar}$
and $\La_{\rm fermion}$ are given, respectively, by:
\begin{align}\label{scalar-PAAP}
{\cal L}_{{\rm scalar}} &= {\rm tr}\Big[i\hbar y_n^\dagger D_t y^n
+i\hbar\hat y^{\dagger n'} D_t\hat y_{n'}
- \frac{\hbar^2}{2m} D_i y_n^\dagger D_i y^n
- \frac{\hbar^2}{2m} D_i \hat y_{n'} D_i \hat y^{\dagger n'}
\nn \\
&~~~~~~~-\frac{\pi\hbar^2}{km}\big(y^n y_n^\dagger y^m y_m^\dagger
-\hat y^{\dagger n'}\hat y_{n'} \hat y^{\dagger m'}  \hat y_{m'}
-y_n^\dagger y^n y_m^\dagger y^m + \hat y_{n'} \hat y^{\dagger n'}
\hat y_{m'}\hat y^{\dagger m'}\big)\Big],
\end{align}
and
\begin{align}\label{fermion-PAAP}
{\cal L}_{{\rm fermion}} &= i\hbar\,{\rm tr}\big(\hat\psi_{+n}
D_t\hat\psi_{+}^n
+\psi_-^{\dagger n'} D_t\psi_{-n'}\big)
-\frac{\hbar^2}{2m}{\rm tr}\Big(D_i\hat\psi_{+n}^{\dagger}D_i\hat\psi_{+}^n
+ D_i\psi_-^{\dagger n'} D_i\psi_{-n'}\Big)
\nn \\
& +\frac{\hbar^2}{2m}{\rm tr}
\Big(\hat\psi_{+n}^{\dagger}\hat F_{12}\hat\psi_{+}^n
-F_{12}\hat\psi_{+n}^{\dagger}\hat\psi_{+}^n
-\psi_-^{\dagger n'}F_{12}\psi_{-n'}
+\hat F_{12}\psi_-^{\dagger n'}\psi_{-n'}\Big)
\nn \\
&+\frac{\pi\hbar^2}{km}{\rm tr}\Big[
\big(y_n^\dagger y^n + \hat y_{n'}\hat y^{\dagger n'}\big)
\big(\hat\psi_+^m\hat\psi_{+m}^\dagger+\psi_-^{\dagger m'}\psi_{-m'}\big)
+\big(y^n y_n^\dagger + \hat y^{\dagger n'}\hat y_{n'}\big)
\big(\hat\psi_{+m}^\dagger\hat\psi_+^m
+ \psi_{-m'}\psi_-^{\dagger m'}\big)
\nn \\
&~~~~~~~~~~~~-2\big(y^{n}y_{m}^\dagger\hat\psi_{+n}^\dagger\hat\psi_+^m
-y^n \hat y_{m'}\hat\psi_{+n}^\dagger\psi_-^{\dagger m'} -\hat
y^{\dagger n'} y_{m}^\dagger\psi_{-n'}\hat\psi_{+}^m +\hat
y^{\dagger n'}\hat y_{m'}\psi_{-n'}\psi_{-}^{\dagger m'}\big)
\nn \\
&~~~~~~~~~~~~-2\big(y_{n}^\dagger y^{m}\hat\psi_{+}^n\hat\psi_{+m}^\dagger
-y_m^\dagger\hat y^{\dagger m'}\hat\psi_{+}^n\psi_{-m'}
-\hat y_{n'} y^{m}\psi_{-}^{\dagger n'}\hat\psi_{+m}^\dagger
+\hat y_{n'}\hat y^{\dagger m'}\psi_{-}^{\dagger n'}
\psi_{-m'}\big)\Big],
\end{align}
while the Chern-Simons part $\La_{CS}$ is still given by \eqref{ABJMcs}.

In deriving the fermionic part of the Lagrangian \eqref{fermion-PAAP},
we used the equations of motion for $\hat\psi_{-n}^{\dagger}$ and
$\psi_+^{\dagger n'}$
\begin{align}
\hat\psi_{-}^n &=-\frac{i\hbar}{2mc}D_-\hat\psi_{+}^n
- \frac{i\hbar}{2mc^2}D_t\hat\psi_{-}^n
=-\frac{i\hbar}{2mc}D_-\hat\psi_{+}^n
+{\cal O}\Big(\frac{1}{m^2c^3}\Big),
\nn \\
\psi_{+n'} &=\frac{i\hbar}{2mc}D_+\psi_{-n'}
- \frac{i\hbar}{2mc^2}D_t\psi_{+n'}=\frac{i\hbar}{2mc}D_+\psi_{-n'}
+{\cal O}\Big(\frac{1}{m^2c^3}\Big),
\end{align}
to eliminate the `heavy modes' $\hat\psi_{-}^n$ and $\psi_{+n'}$, and
$D_\pm = D_1 \pm i D_2$.

\subsection{8 kinematical supercharges}
The action \eqref{PAAP-action} for nonrelativistic PAAP limit is
invariant with respect to supersymmetry transformations.
In Ref.~\cite{Nakayama:2009cz}, eight supersymmetry transformations were found.
This supersymmetry is all kinematical and given by
\begin{align}\label{8susy}
&\delta y^n = -\eta_+^{nn'}\psi_{-n'},\qquad
\delta y_n^\dagger = \eta_{-nn'}\psi_-^{\dagger n'},
\nn \\
&\delta \hat y_{n'} = \eta_{-nn'}\hat\psi_+^n,~~~~~~~~
\delta \hat y^{\dagger n'} = -\eta_{+}^{nn'}\hat\psi_{+n}^\dagger,
\nn \\
&\delta\hat\psi_+^n = -\eta_+^{nn'}\hat y_{n'},~~~~~~~\,
\delta\hat\psi_{+n}^\dagger = -\eta_{-nn'}\hat y^{\dagger n'},
\nn \\
&\delta\psi_{-n'} =\eta_{-nn'} y^n, ~~~~~~~
\delta\psi_-^{\dagger n'} = \eta_+^{nn'} y_n^\dagger,
\nn \\
&\delta A_t = -\frac{\pi \hbar}{km}\big(\hat y^{\dagger n'}\eta_{-nn'}
\hat\psi_+^n +\hat\psi_{+n}^\dagger\eta_+^{nn'}\hat y_{n'}-\eta_+^{nn'}
\psi_{-n'}y_n^\dagger - y^n\psi_-^{\dagger n'}\eta_{-nn'}\big),
\nn \\
&\delta \hat A_t = -\frac{\pi \hbar}{km}\big(\eta_{-nn'}\hat\psi_+^n
\hat y^{\dagger n'} + \hat y_{n'}\hat\psi_{+n}^\dagger\eta_+^{nn'}
-y_n^\dagger\eta_+^{nn'}\psi_{-n'} - \psi_-^{\dagger n'}
\eta_{-nn'}y^n\big),
\nn \\
&\delta A_{\pm} =\delta\hat A_{\pm} = 0,
\end{align}
where $A_{\pm}\equiv A_1\pm i A_2$ and $\hat A_{\pm}\equiv
\hat A_1\pm i \hat A_2$ and the one-component spinor parameters
$\eta_{\pm}$ satisfy the following conditions,
\begin{align}\label{8susypa}
\eta_{\pm nn'} = \frac{1}{2}\epsilon_{nm}\epsilon_{n'm'} \eta_{\pm}^{mm'},
\quad \eta_+^{nn'} = -\eta_+^{n'n}, \quad \eta_{-nn'} = -\eta_{-n'n}
\end{align}
with $\epsilon_{12}=\epsilon_{3'4'}=1$.

\subsection{Superalgebra}

Let us consider the superalgebra of symmetries of the action
\eqref{PAAP-action}. In order to do this we rewrite the action
\eqref{PAAP-action} in a form more appropriate for canonical analysis,
\begin{align}
\La_{\rm PAAP}=&i\hbar\tr\Big(y^\dagger_n\partial_ty^n +{\hat
y}^{\dagger n'}\partial_t{\hat y}_{n'} +{\hat
\psi}^\dagger_{+n}\partial_t{\hat \psi}^n_+ +\psi^{\dagger
n'}_-\partial_t\psi_{-n'} \Big)-\hbar^2{\cal H}
\nn \\
&+\frac{k\hbar}{2\pi}\tr\Big(A_2\partial_tA_1-{\hat
A}_2\partial_t{\hat A}_1
+A_t{\cal G}-{\hat A}_t{\hat
{\cal G}}\Big), \label{action}
\end{align}
where the Hamiltonian density is given by
\begin{align}
{\cal H}&=\frac1{2m}\tr\Big(D_iy^\dagger_nD_iy^n +D_i{\hat
y}^{\dagger n'}D_i{\hat y}_{n'}
+D_i{\hat\psi}^\dagger_{+n}D_i{\hat\psi}^n_++D_i\psi^{\dagger
n'}_-D_i\psi_{-n'}
\nn \\
&~~~~~~~~~~~~~\,-{\hat\psi}^\dagger_{+n}\hat F_{12}\hat\psi^n_+
+F_{12}\hat\psi^\dagger_{+n}\hat\psi^n_+ +\psi^{\dagger
n'}_-F_{12}\psi_{-n'}-\hat F_{12}\psi^{\dagger n'}_-\psi_{-n'}\Big)
\nn \\
&+\frac{\pi}{km}\tr\Big[y^ny^\dagger_ny^my^\dagger_m -\hat
y^{\dagger n'}\hat y_{n'}\hat y^{\dagger m'}\hat y_{m'}
-y^\dagger_ny^ny^\dagger_my^m +\hat y_{n'}\hat y^{\dagger n'}\hat
y_{m'}\hat y^{\dagger m'}
\nn \\
&~~~~~~~~~~~~\,-\big(y_n^\dagger y^n + \hat y_{n'}\hat y^{\dagger
n'}\big) \big(\hat\psi_+^m\hat\psi_{+m}^\dagger+\psi_-^{\dagger
m'}\psi_{-m'}\big) -\big(y^n y_n^\dagger + \hat y^{\dagger n'}\hat
y_{n'}\big) \big(\hat\psi_{+m}^\dagger\hat\psi_+^m +
\psi_{-m'}\psi_-^{\dagger m'}\big)
\nn \\
&~~~~~~~~~~~~\,+2\big(y^{n}y_{m}^\dagger\hat\psi_{+n}^\dagger\hat\psi_+^m
-y^n \hat y_{m'}\hat\psi_{+n}^\dagger\psi_-^{\dagger m'} -\hat
y^{\dagger n'} y_{m}^\dagger\psi_{-n'}\hat\psi_{+}^m +\hat
y^{\dagger n'}\hat y_{m'}\psi_{-n'}\psi_{-}^{\dagger m'}\big)
\nn \\
&~~~~~~~~~~~~\,+2\big(y_{n}^\dagger
y^{m}\hat\psi_{+}^n\hat\psi_{+m}^\dagger -y_m^\dagger\hat y^{\dagger
m'}\hat\psi_{+}^n\psi_{-m'} -\hat y_{n'} y^{m}\psi_{-}^{\dagger
n'}\hat\psi_{+m}^\dagger +\hat y_{n'}\hat y^{\dagger
m'}\psi_{-}^{\dagger n'} \psi_{-m'}\big)\Big],
\end{align}
and the Gauss law constraints are
\begin{align}
\cal G&=F_{12}-\frac{2\pi}{k}\big(y^ny^\dagger_n-\hat y^{\dagger
n'}\hat y_{n'}
-\hat\psi^\dagger_{+n}\hat\psi^n_+-\psi_{-n'}\psi^{\dagger
n'}_-\big),
\nn \\
\hat{\cal G}&=\hat F_{12}-\frac{2\pi}{k}\big(y^\dagger_ny^n -\hat
y_{n'}\hat y^{\dagger n'}
+\hat\psi^n_+\hat\psi^\dagger_{+n}+\psi^{\dagger n'}_-\psi_{-n'}\big).
\end{align}

In the following subsections, we will first derive the superalgebra
associated with the supersymmetry transformation given in (\ref{8susy}).
The analysis parallels to that of the PPPP case~\cite{Nakayama:2009cz}.
The supersymmetry coming from the monopole operator will be studied in the next section.
The canonical commutation relations  are given from (\ref{action}):
\begin{align}
\Big[A_+\big(x\big)^a_b,A_-\big(y\big)^c_d\Big]
&=\frac{4\pi}{k}\delta^a_d\delta^c_b\delta^2\big(x-y\big),~~~~~\,
\Big[\hat A_+\big(x\big)^{\hat
a}_{\hat b},\hat
A_-\big(y\big)^{\hat c}_{\hat
d}\Big]
=-\frac{4\pi}{k}\delta^{\hat
a}_{\hat d}\delta^{\hat
c}_{\hat
b}\delta^2\big(x-y\big),
\nn \\
\Big[y^n\big(x\big)^a_{\hat
a},y^\dagger_m\big(y\big)^{\hat
b}_b\Big]
&=\delta^n_m\delta^a_b\delta^{\hat
b}_{\hat
a}\delta^2\big(x-y\big),~~~~~\,
\Big[\hat
y_{n'}\big(x\big)^{\hat
a}_a,\hat y^{\dagger
m'}\big(y\big)^b_{\hat b}\Big]
=\delta^{m'}_{n'}\delta^b_a\delta^{\hat
a}_{\hat
b}\delta^2\big(x-y\big),
\nn \\
\Big\{\hat\psi^n_+\big(x\big)^{\hat
a}_a,\hat\psi^\dagger_{+m}\big(y\big)^b_{\hat
b}\Big\}
&=\delta^n_m\delta^b_a\delta^{\hat
a}_{\hat
b}\delta^2\big(x-y\big),~~
\Big\{\psi_{-n'}\big(x\big)^a_{\hat
a},\psi^{\dagger
m'}_-\big(y\big)^{\hat
b}_b\Big\}
=\delta^{m'}_{n'}\delta^a_b\delta^{\hat
b}_{\hat
a}\delta^2\big(x-y\big).
\end{align}

\subsubsection{Bosonic Schr\"{o}dinger algebra}

Let us first consider the bosonic subalgebra. Applying Noether techniques
we find that the bosonic symmetry algebra consists of the following
conserved charges:
\begin{itemize}
\item Hamiltonian
\begin{align}
H=\int d^2x~\cal H.
\end{align}
\item Linear momentum
\begin{align}
P_i=\int d^2x~{\cal P}_i,
\end{align}
where the momentum density is given by
\begin{align}
{\cal P}_i&=-\frac{i}{2}\tr\Big(y^\dagger_nD_iy^n-D_iy^\dagger_ny^n +{\hat
y}^{\dagger n'}D_i{\hat y}_{n'}-D_i{\hat y}^{\dagger n'}{\hat
y}_{n'}
\nn \\
&~~~~~~~~~~~~\,+\hat\psi^\dagger_{+n}D_i\hat\psi^n_+
-D_i\hat\psi^\dagger_{+n}\hat\psi^n_+ +\psi^{\dagger n'}_-
D_i\psi_{-n'}-D_i\psi^{\dagger n'}_-\psi_{-n'}\Big).
\end{align}
\item Angular momentum
\begin{align}
J=\int d^2x~\epsilon_{ij}x_i{\cal P}_j+\Sigma,
\end{align}
where the spin $\Sigma$ is given by
\begin{align}\label{Sig}
\Sigma=\frac12\int d^2x~\tr\Big(\psi^{\dagger n'}_-\psi_{-n'}
-\hat\psi^\dagger_{+n}\hat\psi^n_+\Big).
\end{align}
\item Total number operator
\begin{align}
{\cal N}=\int d^2x~n\big(x\big)\label{numch}
\end{align}
with number density given by
\begin{align}
n\big(x\big)=\tr\Big(y^\dagger_ny^n+{\hat y}^{\dagger n'}{\hat
y}_{n'} +\hat\psi^\dagger_{+n}\hat\psi^n_++\psi^{\dagger
n'}_-\psi_{-n'}\Big).
\end{align}
\item Galilean boost
\begin{align}
G_i=-tP_i+m\int d^2x~x_in\big(x\big).
\end{align}
\item Dilatation
\begin{align}
D=2tH-\int d^2x~x_i{\cal P}_i. \end{align}
\item Special conformal transformation
\begin{align}
K=-t^2H+tD+\frac{m}{2}\int d^2x~x^2n\big(x\big).
\end{align}
\end{itemize}

These operators are the subjects to the following algebraic relations
defining the bosonic Schr\"{o}dinger algebra~\cite{Jackiw:1990mb}:
\begin{align}
i\Big[D,H\Big]&=2H,~~~ i\Big[D,K\Big]=-2K,~~~ i\Big[K,H\Big]=D,
\nn \\
i\Big[H,G_i\Big]&=P_i,~~~~\, i\Big[K,P_i\Big]=-G_i, ~~~
i\Big[P_i,G_i\Big]=\delta_{ij}m\cal N,
\nn \\
i\Big[D,P_i\Big]&=P_i,~~~~ i\Big[D,G_i\Big]=-G_i, ~~~~\,
i\Big[J,P_i\Big]=-\epsilon_{ij}P_j,~~~~
i\Big[J,G_i\Big]=-\epsilon_{ij}G_j. \label{sch}
\end{align}
In the above algebra $H, D, K$ are the generators of the conformal subgroup
$SO(2,1)$ and the number operator ${\cal N}$ appears as a
central term.

\subsubsection{Kinematical superalgebra}\label{kinsalg}
The bosonic algebra can be enlarged to superalgebra by adding the 8 kinematic
fermionic charges generating the transformations (\ref{8susy}),
\begin{align}
q^{nn'}_+&=i\int d^2x~\tr\Big(\psi^{\dagger n'}_-y^n
-\hat{y}^{\dagger n'}\hat\psi^n_+\Big),
\nn \\
q_{-nn'}&=i\int d^2x~\tr\Big(\hat\psi^\dagger_{+n}\hat{y}_{n'}
-y^\dagger_n\psi_{-n'} \Big),
\end{align}
 as well as
$SU(2)_{{\rm L}}$ and $SU(2)_{{\rm R}}$
$R$-symmetry generators
\begin{align}
R^n_m &=-\int
d^2x~\tr\Big[y^\dagger_my^n+\hat\psi^\dagger_{+m}\hat\psi^n_+
-\frac12\delta^n_m\big(y^\dagger_ly^l+\hat\psi^\dagger_{+l}
\hat\psi^l_+\big)\Big],\nn\\
R^{n'}_{m'}&=\int d^2x~\tr\Big[\hat{y}^{\dagger n'}\hat{y}_{m'}
+\psi^{\dagger n'}_-\psi_{-m'}
-\frac12\delta^{n'}_{m'}\big(\hat{y}^{\dagger l'}\hat{y}_{l'}
+\psi^{\dagger l'}_-\psi_{-l'}\big) \Big].
\end{align}
They  rotate the fields and the
generators with $SU(2)_{{\rm L}}$ and
$SU(2)_{{\rm R}}$ indices. Therefore they satisfy the standard su(2) algebra.
The supercharges satisfy the following commutation relations
\begin{align}
\Big\{q^{nn'}_+,q_{-mm'}\Big\}=\frac12\delta^{n'}_{m'}\delta^n_m{\cal
N}
+\delta^n_mR^{n'}_{m'}-\delta^{n'}_{m'}R^n_m,\nn\\
\Big[J,q^{nn'}_+\Big]=\frac12q^{nn'}_+, \qquad
\Big[J,q_{-nn'}\Big]=-\frac12q_{-nn'}.\label{kin}
\end{align}
Along with Schr\"{o}dinger algebra (\ref{sch}), the commutation relations
(\ref{kin}) describe a {\it kinematical superalgebra}.

In the case of PPPP, there exists an  $U(1)_R$ symmetry in
the non-relativistic limit coming from the reduction of $SU(4)$
$R$-symmetry to $SU(2)_{{\rm L}}\times SU(2)_{{\rm R}}\times U(1)_R$ symmetry. We
will see that in the PAAP case also, there is a similar $U(1)_R$
symmetry connecting the particle and antiparticle sectors.

\section{SUSY Enhancement with Monopole Operators in Nonrelativistic Limit}

In this section we find the additional supercharges and the corresponding
superalgebra in the PAAP limit. We emphasize the fundamental role of
monopole operators in the supersymmetry enhancement for both
relativistic and nonrelativistic superconformal field theories.

\subsection{SUSY enhancement and monopole operators}\label{mono}
For the Chern-Simons level $k=1,2$ the ABJM theory was shown to be a
subject to the additional hidden $\mathcal{N}=2$
supersymmetry~\cite{Gustavsson:2009pm,Kwon:2009ar}.
The additional supersymmetry transformations involve
the monopole operators.
In this subsection we investigate the role of monopole operators
in supersymmetry enhancement of (mass deformed) ABJM theory and
its nonrelativistic limit.

The ABJM theory described by the action (\ref{ABJMact})
contains complex representations for matter fields
and it has SU(4)$\times$U(1) global symmetry corresponding to
the transverse space to M2-branes.
To find the hidden symmetry in the transverse space we should use
both the {\bf 4} and ${\bf \bar 4}$-complex representations simultaneously.
However, in order to include both of them in the same supersymmetry
transformation, we have to modify one of the representations by the help
of a monopole operator.

Attaching the monopole operator to a matter field changes
the gauge transformation rule of the field without changing the
global symmetry properties.
For instance, a bi-fundamental scalar field $Y^A$ in {\bf 4}-representation
can be converted to an anti-bi-fundamental one $\tilde Y^A$
in the same {\bf 4}-complex representation as follows,
\begin{align}
\tilde Y^{A\hat a}_{~~~a} = \bar\tau^{\hat a\hat b}_{ab} Y^{Ab}_{~~~\hat b},
\end{align}
where $\bar\tau^{\hat a\hat b}_{ab}$ is the monopole operator related
to the supersymmetry enhancement. For the U(2)$\times$U(2) case,
the explicit form of the monopole operator is given by~\cite{Kwon:2009ar}
\begin{align}
\bar\tau^{\hat a\hat b}_{ab} = e^{-2i\int_x^\infty A_\mu^-dx^\mu}
\epsilon^{\hat a\hat b}\epsilon_{ab},
\end{align}
where $\epsilon_{ab}$ and $\epsilon^{\hat a\hat b}$ are invariant antisymmetric
tensors for left and right SU(2) parts of the gauge group respectively.

Attaching a monopole operator to a
local field also does not change the local nature of the resulting
composite field for $k=1,2$. In this case, we can introduce
in supersymmetry transformations new local fields,
which do not exist in the original ABJM theory.
This fact was demonstrated for the abelian ABJM
theory~\cite{Gustavsson:2009pm}.

Therefore the enhanced supersymmetry transformation rule takes the
following form~\cite{Kwon:2009ar}
\begin{align}\label{deltaY}
\delta Y^A = i\omega^{AB}\Psi_B + i\tilde\varepsilon\tilde\Psi^{\dagger A},
\end{align}
where $\tilde\Psi^{\dagger Aa}_{~~~~\hat a} =
\tau^{ab}_{\hat a\hat b}\Psi^{\dagger A\hat b}_{~~~~b}$ is in the
bi-fundamental representation and $\tilde\varepsilon$ is a complex spinor.
As we see in (\ref{deltaY}), we have to use the
{\bf 4} and ${\bf\bar 4}$-representations simultaneously in order to obtain
the additional ${\cal N}=2$ supersymmetry. Due to the presence of the
monopole operator both sides of expression in (\ref{deltaY}) can be in
the bi-fundamental representations of the gauge group.

Now let us consider the supersymmetric properties and the role of
monopole operator in nonrelativistic limits.
Even though it is possible to obtain the supersymmetry transformation rules
in nonrelativistic limit by reduction from those of relativistic theory,
we use different strategy. Instead we analyze full supersymmetry of the
nonrelativistic action.
However, taking into account the nature of the monopole operators,
we can roughly estimate the supersymmetric properties for a
given choice of particle and antiparticle sectors.

In PPPP limit, (\ref{deltaY}) is reduced to
\begin{align}\label{dynnp}
\delta y^n &= i\sqrt{\frac{2mc}{\hbar}}\omega^{nm}\psi_m
+i\sqrt{\frac{2mc}{\hbar}}\omega^{nm'}\psi_{m'}
+i\sqrt{\frac{2mc}{\hbar}} e^{2imc^2t/\hbar}\tilde\varepsilon
\tilde\psi^{\dagger n},
\nn \\
\delta y^{n'} &= i\sqrt{\frac{2mc}{\hbar}}\omega^{n'm}\psi_m
+i\sqrt{\frac{2mc}{\hbar}}\omega^{n'm'}\psi_{m'}
+i\sqrt{\frac{2mc}{\hbar}} e^{2imc^2t/\hbar}\tilde\varepsilon
\tilde\psi^{\dagger n'},
\end{align}
where $\tilde\psi^{\dagger Aa}_{~~~~\hat a} = \tau^{ab}_{\hat a\hat b}
\psi^{\dagger A\hat b}_{~~~~b}$.
Dropping the fast oscillating terms in the nonrelativistic limit
$c\to\infty$ in (\ref{dynnp}),
we see that the
supersymmetry parameter $\tilde\varepsilon$ in (\ref{dynnp}) does not exist
in the limit.
This means that the additional ${\cal N}=2$ supersymmetry with the monopole
operator is broken in the PPPP limit.
Then the PPPP limit has only 12 supercharges mentioned in
Refs.~\cite{Nakayama:2009cz,Lee:2009mm},
which are inherited from the original ${\cal N}=6$ supersymmetry
in the ABJM theory.
There are also 2 conformal supercharges which are new in nonrelativistic limit.

On the other hand, for the PAAP limit for $k=1,2$, (\ref{deltaY}) becomes
\begin{align}\label{dynnp2}
\delta y^n &= i\sqrt{\frac{2mc}{\hbar}}e^{2imc^2t/\hbar}\omega^{nm}
\sigma_2\hat\psi^{\dagger}_m
+i\sqrt{\frac{2mc}{\hbar}}\omega^{nm'}\psi_{m'}
+i\sqrt{\frac{2mc}{\hbar}}\tilde\varepsilon
\tilde{\hat \psi}^{n},
\nn \\
\delta \hat y^{\dagger n'} &= i\sqrt{\frac{2mc}{\hbar}}\omega^{n'm}\sigma_2
\hat\psi^{\dagger }_m
+i\sqrt{\frac{2mc}{\hbar}}e^{-2imc^2t/\hbar}\omega^{n'm'}\psi_{m'}
+i\sqrt{\frac{2mc}{\hbar}}\tilde\varepsilon
\tilde\psi^{\dagger n'},
\end{align}
where $\tilde{\hat \psi}^{na}_{~~\hat a} = \tau^{ab}_{\hat a\hat b}
{\hat \psi}^{n \hat b}_{~~b}$.
Similarly to the PPPP case, $\omega^{nm}$ (dual of $\omega^{n'm'}$)
should vanish in this limit. Instead, $\tilde\varepsilon$
survives. So ${\cal N}=4$ part having supersymmetry parameter $\omega^{nm'}$
in the original ${\cal N}=6$ of the ABJM theory,
the additional ${\cal N}=2$ part with monopole operator,
and the emergent 2 conformal supercharges are supercharges in the PAAP limit.
As a result, the total number of supercharges in the PAAP limit is the same as
in the PPPP case.

\subsection{SUSY enhancement in the PAAP limit}\label{susypaap}

Let us go back to the nonrelativistic action in the PAAP limit.
In fact, the supersymmetry of the PAAP action \eqref{PAAP-action} is much richer
than the eight kinematical supersymmetry transformations given
in (\ref{8susy}).
In addition, we have two more kinematical
supercharges constructed from the monopole operators,
\begin{align}
&\delta y^n = -\check\xi_-\tau\cdot\hat\psi_+^n,~~~~~~~~\,
\delta y_n^\dagger =\check\xi_-^\dagger\bar\tau\cdot\hat\psi_{+n}^\dagger,
\nn \\
&\delta \hat y_{n'} = -\check\xi_-^\dagger\bar\tau\cdot\psi_{-n'},\qquad
\delta \hat y^{\dagger n'} = \check\xi_-\tau\cdot\psi^{\dagger n'},
\nn \\
&\delta\hat\psi_+^n = \check\xi_-\bar\tau\cdot y^n,~~~~~~~~~~~
\delta\hat\psi_{+n}^\dagger = \check\xi\tau\cdot y_n^\dagger,
\nn \\
&\delta\psi_{-n'} =\check\xi_-\tau\cdot\hat y_{n'},~~~~~~~~\,
\delta\psi_-^{\dagger n'} = \check\xi_-^{\dagger}\bar\tau\cdot
\hat y^{\dagger n'},
\nn \\
&\delta A_t = \frac{\pi \hbar}{km}\big(\tau\cdot y_n^\dagger\check\xi_-
\hat\psi_+^n -\check\xi_-^\dagger\hat\psi_{+n}^\dagger\bar\tau\cdot y^n
+\check\xi_-^\dagger\psi_{-n'}\bar\tau\cdot\hat y^{\dagger n'}
-\tau\cdot\hat y_{n'}\check\xi_-\psi_-^{\dagger n'}\big),
\nn \\
&\delta \hat A_t =\frac{\pi \hbar}{km}\big(\check\xi_-\hat\psi_-^n
\tau\cdot y_n^\dagger-\bar\tau\cdot
y^n\check\xi_-^\dagger\hat\psi_{+n}^\dagger,
+\bar\tau\cdot\hat y^{\dagger n'}\check\xi_-^\dagger\psi_{-n'}
-\check\xi_-\psi_-^{\dagger n'}\tau\cdot\hat y_{n'}\big),
\nn \\
&\delta A_{\pm} =\delta\hat A_{\pm} = 0,\label{kinetic}
\end{align}
where $\check\xi_-$ is a one-component complex spinor parameter.
The action of monopole operators on the fields is expressed as
\begin{align} \nn
(\tau\cdot y^\dagger)^a_{~\hat a} = \tau_{\hat a\hat b}^{ab}
{y^\dagger}^{\hat b}_{~b}, \qquad
(\bar\tau\cdot y)^{\hat a}_{~ a} = \bar\tau^{\hat a\hat b}_{ab}
{y}^{b}_{~\hat b},
\end{align}
where $y$ and $y^\dagger$ are in the bi-fundamental and anti-bi-fundamental
representations respectively.

In addition there are two dynamical enhanced supersymmetries which are also
constructed from the monopole operators,
\begin{align}
&\delta y^n = \frac{i}{2m}\check\xi_+\tau\cdot D_-\hat\psi_+^n,~~~~~~~~~~\,
\delta y_n^\dagger = \frac{i}{2m}\check\xi_+^\dagger\bar\tau\cdot D_+
\hat\psi_{+n}^\dagger,
\nn \\
&\delta y_{n'} = \frac{i}{2m}\check\xi_+^\dagger\bar\tau\cdot D_+
\psi_{-n'},~~~~~~~~
\delta y^{\dagger n'} =\frac{i}{2m}\check\xi_+ \tau\cdot D_-
\psi_-^{\dagger n'},
\nn \\
&\delta\hat\psi_+^n = -\frac{i}{2m}\check\xi_+^\dagger\bar\tau\cdot D_+y^n,
~~~~~~~~\,
\delta\hat\psi_{+n}^\dagger =\frac{i}{2m}\check\xi_+\tau\cdot D_- y_n^\dagger,
\nn \\
&\delta\psi_{-n'} =-\frac{i}{2m}\check\xi_+\tau\cdot D_- \hat y_{n'}, \qquad
\delta\psi_-^{\dagger n'} = \frac{i}{2m}\check\xi_+^\dagger\bar\tau\cdot
D_+\hat y^{\dagger n'},
\nn \\
&\delta A_t = \frac{i\pi\hbar}{2km^2}\big(\tau\cdot D_-\hat\psi_+^n
\check\xi_+ y_n^\dagger + y^n\bar\tau\cdot D_+\hat\psi_{+n}^\dagger
\check\xi_+ + \tau\cdot D_-\psi_-^{\dagger n'}\check\xi_+\hat y_{n'}
+\hat y^{\dagger n'}\bar\tau\cdot D_+\psi_{-n'}\check\xi_+^\dagger\big),
\nn \\
&\delta A_+ = \frac{2\pi}{km}\big(
\tau\cdot\hat\psi_+^n\check\xi_+ y_n^\dagger
-\tau\cdot\psi_-^{\dagger n'}\check\xi_+\hat y_{n'}\big),
\nn \\
&\delta A_- = \frac{2\pi}{km}\big(\hat y^{\dagger n'}\bar\tau\cdot
\psi_{-a'}\check\xi_+^\dagger - y^n \bar\tau\cdot\hat\psi_{+n}^\dagger
\check\xi_+^\dagger\big),
\nn \\
&\delta \hat A_t = \frac{i\pi\hbar}{2km^2}\big(
\bar\tau\cdot D_+\hat\psi_{+n}^\dagger \check\xi_+^\dagger y^n
+ y_n^\dagger\tau\cdot D_-\hat\psi_{+}^n \check\xi
+ \bar\tau\cdot D_+\psi_{-n'}\check\xi_+^\dagger\hat y^{\dagger n'}
+\hat y_{n'}\tau\cdot D_-\psi_{-}^{\dagger n'}\check\xi_+\big),
\nn \\
&\delta \hat A_+ = \frac{2\pi}{km}\big(
y_n^\dagger\tau\cdot\hat\psi_+^n\check\xi_+
-\hat y_{n'}\tau\cdot\psi_-^{\dagger n'}\check\xi_+\big),
\nn \\
&\delta \hat A_- = \frac{2\pi}{km}\big(
\bar\tau\cdot\psi_{-n'}\check\xi_+^\dagger \hat y^{\dagger n'} -
\bar\tau\cdot\hat\psi_{+n}^\dagger\check\xi_+^\dagger
y^n\big),\label{dynamical}
\end{align}
where $\check\xi_+$ is a one-component spinor parameter.

\subsection{Dynamical superalgebra}
In this section, we construct the supercharges coming from the transformations
given in the previous subsection \ref{susypaap}.
Together with the $U(1)_R$ $R$-symmetry
mentioned in the subsubsection \ref{kinsalg} (its expression is given below in
(\ref{rr})) and the bosonic
 Schr\"{o}dinger algebra (\ref{sch}), they form
a {\it dynamical superalgebra}.

First, the 2 kinematical supercharges related with the
transformation of (\ref{kinetic}) are given by
\begin{align}
q_+&=-i\int d^2x~\tr\Big(y^\dagger_n\tau\cdot\hat\psi^n_+
+\psi^{\dagger n'}_-\tau\cdot\hat{y}_{n'}\Big),
\nn \\
q_-&=i\int d^2x~\tr\Big(\hat{y}^{\dagger n'}
\bar\tau\cdot\psi_{-n'}+\hat\psi^\dagger_{+n}\bar\tau\cdot y^n\Big),
\end{align}
where $q_-=q^\dagger_+$. The 2 dynamical supercharges associated with the
transformations given in (\ref{dynamical}) are obtained as
\begin{align}
Q_+&=-\frac1{2m}\int d^2x~ \tr\Big(y^\dagger_n\tau\cdot
D_-\hat\psi^n_+ +\psi^{\dagger n'}_-\tau\cdot D_-\hat{y}_{n'}\Big),
\nn \\
Q_-&=\frac1{2m}\int d^2x~\tr\Big(\hat{y}^{\dagger n'}\bar\tau\cdot
D_+\psi_{-n'}+\hat\psi^\dagger_{+n}\bar\tau\cdot D_+y^n\Big),
\end{align}
where $Q_-=Q^\dagger_+$. In addition, the 2 conformal supercharges
which satisfy $i[K,Q_{\pm}]=S_{\pm}$ are given by
\begin{align}
S_+&=tQ_++\frac{i}2\int d^2x~\big(x_1-ix_2\big)
\tr\Big(y^\dagger_n\tau\cdot\hat\psi^n_+ +\psi^{\dagger
n'}_-\tau\cdot\hat{y}_{n'}\Big),
\nn \\
S_-&=tQ_--\frac{i}2\int d^2x~\big(x_1+ix_2\big)
\tr\Big(\hat{y}^{\dagger n'}
\bar\tau\cdot\psi_{-n'}+\hat\psi^\dagger_{+n}\bar\tau\cdot y^n\Big),
\end{align}
where $S_-=S^\dagger_+$. The $U(1)_R$ $R$-symmetry generator is
written as
\begin{align}
R=-\frac12\int d^2x~\tr\Big(y^\dagger_ny^n+\psi^{\dagger
n'}_-\psi_{-n'} -\hat{y}^{\dagger
n'}\hat{y}_{n'}-\hat\psi^\dagger_{+n}\hat\psi^n_+\Big).\label{rr}
\end{align}
Note that this generator represents difference between the number of
particles and that of antiparticles. Recall that the total number operator
(\ref{numch}) is also conserved. Therefore in our nonrelativistic limit
particle and antiparticle numbers are conserved separately.

Putting everything together, we obtain
the following {\it dynamical superalgebra} whose (anti)commutation relations
are given by\footnote{The
sub-index $+(-)$ in the supercharges denotes spin up(down) state.}:
\begin{align}
\Big\{Q_+,Q_-\Big\}&=\frac{1}{2m}H,~~~\Big\{S_+,S_-\Big\}=\frac1{2m}K,
~~~\,\Big\{q_+,q_-\Big\}={\cal N},
\nn \\
\Big\{q_+,Q_-\Big\}&=\frac1{2m}P_+,~~\Big\{q_-,Q_+\Big\}=\frac1{2m}P_-,
~~\Big\{q_+,S_-\Big\}=-\frac1{2m}G_+,~~\Big\{q_-,S_+\Big\}=-\frac1{2m}G_-,
\nn \\
\Big\{Q_+,S_-\Big\}&=\frac1{4m}\Big[D+i\big(J-\frac32\tilde{R}\big)\Big],
~~~~~~~~~~\,\Big\{Q_-,S_+\Big\}=\frac1{4m}
\Big[D-i\big(J-\frac32\tilde{R}\big)\Big],
\label{dynamic1}
\end{align}
and
\begin{align}
i\Big[K,Q_{\pm}\Big]&=S_{\pm},~~\qquad i\Big[G_+,Q_+\Big]=-q_+,
~~~~~i\Big[G_-,Q_-\Big]=-q_-,
\nn \\
i\Big[H,S_{\pm}\Big]&=-Q_{\pm},\qquad i\Big[P_+,S_+\Big]=-q_+,
\qquad i\Big[P_-,S_-\Big]=-q_-,
\nn \\
i\Big[D,Q_{\pm}\Big]&=Q_{\pm},~~~~~~~~~\,i\Big[D,S_{\pm}\Big]=-S_{\pm},
~~~~~~~~\,\Big[D,q_{\pm}\Big]=0,
\nn \\
\Big[J,Q_{\pm}\Big]&=\mp\frac12Q_{\pm},
~\qquad\Big[J,S_{\pm}\Big]=\mp\frac12S_{\pm},
~~~~~~~\,\Big[J,q_{\pm}\Big]=\pm\frac12q_{\pm},\label{dynamic2}
\end{align}
where we introduce $\tilde{R}=\frac23\big(2R+\Sigma\big)$, $P_\pm=P_1\pm iP_2$,
and $G_\pm=G_1\pm iG_2$. The commutation relations for $R$ are
\begin{align}
\Big[R,Q_{\pm}\Big]&=\mp Q_{\pm}, \qquad\Big[R,S_{\pm}\Big]=\mp
S_{\pm}, \qquad\Big[R,q_{\pm}\Big]=\mp q_{\pm},
\nn \\
\Big[R,q^{nn'}_+\Big]&=0, ~~~~~~~~\,\Big[R,q_{-nn'}\Big]=0.
\label{dynamic3}
\end{align}
Then the additional commutation relations can be computed as
\begin{align}
\Big[\tilde{R},Q_{\pm}\Big]&=\mp Q_{\pm},
\qquad\Big[\tilde{R},S_{\pm}\Big]=\mp S_{\pm},
\qquad\Big[\tilde{R},q_{\pm}\Big]=\mp q_{\pm},
\nn \\
\Big[\tilde{R},q^{nn'}_+\Big]&=\frac13q^{nn'}_+,
~~~\Big[\tilde{R},q_{-nn'}\Big]=-\frac13q_{-nn'}.
\end{align}

The   super Schr\"{o}dinger
algebra~\cite{Leblanc:1992wu, Duval:1993hs,Sakaguchi:2008rx}
consists of bosonic Schr\"{o}dinger algebra (\ref{sch}),
{\it kinematical superalgebra}
(\ref{kin}), and the {\it dynamical superalgebra}
(\ref{dynamic1}),(\ref{dynamic2}),(\ref{dynamic3}).
This super Schr\"{o}dinger algebra is isomorphic to the one
in the PPPP limit~\cite{Nakayama:2009cz,Lee:2009mm}.

\section{Conclusion}

In this work we considered  nonrelativistic limits of the ABJM theory.
In particular, we focused on the PAAP limit containing both particles
and antiparticles.
This case was considered previously in the literature \cite{Nakayama:2009cz}.
The known result was that in this limit there are only eight kinematical supercharges.
This is in contrast to the case of PPPP limit containing only
particles~\cite{Nakayama:2009cz,Lee:2009mm}.
In PPPP limit one has fourteen independent supercharges:
In addition to the eight kinematical supercharges there are two dynamical,
two more kinematical as well as two conformal supercharges.

In this work we revisited the PAAP limit and among others we found
that if we extend our considerations to supersymmetry transformations
involving monopole operators, then there are six additional supercharges
besides the basic eight kinematical supercharges found
in \cite{Nakayama:2009cz}.
These additional supercharges consist of two dynamical,
two kinematical, and two conformal ones.

At the same time
we analyzed the PPPP limit as well and found that
there is no additional supersymmetry enhancement due to monopole operator
in this limit. It appears that supersymmetry in both PPPP and PAAP
limit match as a number as well as a structure: ten kinematical,
two dynamical and two conformal supercharges. In fact, both supersymmetry algebras
are isomorphic to the super Schr\"{o}dinger algebra with fourteen supercharges.

These results give some insight into the r\^{o}le of the monopole
operator in ABJM theory.
First of all the existence of additional supercharges in the
nonrelativistic PAAP limit is possible because the monopole
operators map antiparticle states into particle states extending
the supersymmetry transformations to one mixing particles with
antiparticles. This also explains why one can not observe any
extended supersymmetries in the PPPP limit where there are only particles.
In the nonrelativistic limit the particles and antiparticles are
explicitly separated and therefore the physical meaning of the monopole
operator which mixes them becomes particularly clear.

The fact that both PPPP
and PAAP limits possess isomorphic supersymmetries suggests the
conjecture that \emph{the super Schr\"{o}dinger algebra with fourteen
supercharges is the maximal unbroken supersymmetry in the
nonrelativistic limit}. Moreover, out of the fourteen supercharges
two of them seem to be new: the conformal supercharges appear due to the
emergence of nonrelativistic conformal symmetry in $c\to\infty$ limit.
Therefore only twelve supersymmetries can be considered as ``genuine''
ones coming from the relativistic theory. This corresponds to $\N=6$
relativistic spinor supercharges from $\N=8$ total supersymmetry
including the monopole enhanced supersymmetry part.
The technical part of this supersymmetry breaking is clear:
For either choice of nonrelativistic limit
there are at least two kinds of supersymmetry transformation which are
producing fields with wrong phase, therefore they should be broken.
However, more fundamental reason for this breaking is not clear to us.
A further study may clarify this issue.

We restricted our detailed analysis to the case of U(2)$\times$U(2)
gauge group, because in this case the monopole operator as well as
the corresponding supersymmetry transformations can be constructed
explicitly. It would be
interesting, however, to apply our analysis to the enhanced supersymmetry of
ABJM theory with U$(N)\times$U$(N)$ gauge symmetry in the
three-algebra approach of Ref.~\cite{Gustavsson:2009pm}.
Furthermore, these results might be generalized to any
nonrelativistic limit of superconformal field
theories~\cite{Nakayama:2008qm,Nakayama:2008qz} which include both
particles and antiparticles, and allow supersymmetries generated by
the monopole operators. These aspects also deserve further investigation.

Another important subject that remains to be considered  is
to explore the M-theory dual of the
observed phenomena.
This should be of particular interest for the AdS dual description of
low dimensional condensed matter systems. Let us note that the present
approach to AdS/CMT correspondence is mainly based on the embedding
of the nonrelativistic system into a higher-dimensional relativistic
one using the light-cone approach.
The approach we followed here is based on the nonrelativistic
reduction of a relativistic system in the same dimension which
seems to be more natural since any condensed matter system is, in fact,
a nonrelativistic limit of a relativistic system.

\subsection*{Acknowledgements}
We would like to thank Shinsuke Kawai for helpful discussions. O.K.,
C.S., and J.S. would like to thank the APCTP for its hospitality,
during the workshop  ``APCTP Focus Program on Current Trends in
String Field Theory'', where part of this work was done.
This work is supported by the National Research Foundation of Korea(NRF) grant
funded by the Korea government(MEST) through the Center for Quantum
Spacetime(CQUeST) of Sogang University with grant number 2005-0049409(P.O.)
and the Korea Research Foundation(KRF) grant funded by the
Korea government(MEST) (No. 2009-0073775) (O.K.).

\providecommand{\href}[2]{#2}\begingroup\raggedright\endgroup

\end{document}